\documentclass[12pt,a4paper]{article}
\usepackage{amssymb}
\usepackage{amsfonts}
\usepackage{graphicx}
\usepackage[hyperindex=true,
          pdfstartview=FitH,
          bookmarksnumbered=true,
          bookmarksopen=true,
          citecolor=blue,
          linkcolor=blue,
          colorlinks=true,
          pdfborder=001,
          unicode]{hyperref}

\begin{document}
\title{Five-dimensional vacuum Einstein spacetimes\\
in C-metric like coordinates}
\author{Wei Xu, Liu Zhao\thanks{Correspondence author} \ and Bin Zhu\thanks{emails: {\it
xuweifuture@mail.nankai.edu.cn}, 
{\it lzhao@nankai.edu.cn} and {\it binzhu7@gmail.com}}\\
School of Physics, Nankai University, \\
Tianjin 300071, P R China}
\date{}
\maketitle
\begin{abstract}
A 5-dimensional Einstein spacetime 
with (non)vanishing cosmological constant is analyzed in detail. 
The metric is in close analogy with the 4-dimensional massless 
uncharged C-metric in many
aspects. The coordinate system, horizons and causal structures, 
relations to standard form of de Sitter, anti de Sitter and Minkowski 
vacua are investigated. After a boost and Kaluza-Klein reduction, we get 
an exact solution of 4-dimensional Einstein-Maxwell-Liouville theory 
which reduces to a solution to Einstein-Liouville theory in the limit of 
zero boost velocity and to that of Einstein-Maxwell-dilaton theory in the 
case of zero cosmological constant.
\end{abstract}

\section{Introduction}
The study of higher-dimensional gravity attracted considerable attentions
in the recent years. The reason is two folded: one reason lies in that string
theory requires higher spacetime dimensions, the other reason is that
studies of gravity in higher dimensions might reveal deeper structures of
general relativity, such as stability issues and classification of singularities 
of spacetime in various dimensions. For recent reviews in the subject, see
\cite{Emparan:2008} and \cite{Obers}.

In \cite{Liu Zhao}, two of the present authors obtained and analyzed in
detail an exact 5-dimensional (5D) C-metric like vacuum solution of the
Einstein equation with non-negative cosmological constant. The C-metric
in 4D has been known for a long time \cite{Witten}. The
original C-metric and subsequent generalizations were shown to admit very
interesting physical features \cite{Emparan:2001, Griffiths, Hong,
Dias:2002, Dias:2003}. Despite the intensive studies on higher dimensional 
gravities, it is a bit strange that no higher dimensional analogue of  
C-metric has been found so far. It was concluded very recently
\cite{Podolsky-talk}\cite{Podolsky-talk2} that there are no 
generalizations of the C-metric in 
the Robinson-Trautman family with black hole 
horizons in higher dimensions. The spacetime
we obtained in \cite{Liu Zhao} does not contain any black hole horizons, 
but it does have
two acceleration horizons and interestingly nontrivial global structure. We
also interpreted the special case of $\Lambda =0$ using exterior geometric
technique following \cite{Frolov}.

In the discussion section of  \cite{Liu Zhao},
another C-metric like vacuum solution was presented but not analyzed in
detail. In this paper, we shall be dealing with this novel metric with
arbitrary value of $\Lambda$. The plan of this paper is as follows. In 
section 2 we
give the metric and its coordinate ranges. In section 3 we study the 
causal structures of the metric in various cases. In section 4 we analyze the 
accelerating nature of the horizons. Section 5 is devoted to the connection 
to the standard dS, AdS and Minkowski spacetimes using exterior 
geometric techniques. Then, in section 6, we give a 4D
interpretation of the metric via Kaluza-Klein reduction, which leads to an 
exact solution for Einstein-Maxwell-Liouville theory just as the metric 
analyzed in \cite{Liu Zhao}.
Finally, in section 7, some concluding remarks are presented.

\section{The metric and coordinate ranges} \label{sec2}

The metric to be studied is
\begin{eqnarray}
 \mathrm{d} s^2 = \frac{1}{\alpha^2(x+y) ^2}\left[-G(y) \mathrm{d} t^2
 + \frac{\mathrm{d} y^2}{G(y) } + \frac{\mathrm{d} x^2}{F(x) } + F(x)
 \left(\frac{\mathrm{d} z^2}{H(z) } + H(z) \mathrm{d} \phi^2
 \right) \right],
 \label{metric1}
\end{eqnarray}
where
\begin{eqnarray}
 F(x) =1-x^2, \quad G(y) =-1-\frac{\Lambda}{6\alpha^2}+y^2,
 \quad H(z) =1-z^2.
\end{eqnarray}
By straightforward calculations it can be seen that this metric is an exact 
solution of the 5D vacuum Einstein equation $R_{MN}-\frac{1}{2}g_{MN}R +\Lambda g_{MN}=0$. So the corresponding spacetime is a 5D vacuum Einstein spacetime for every constant value of the cosmological constant $\Lambda$.

Obviously, the function $F(x) $ and $H(z) $ both have two zeros
\begin{eqnarray*}
 x=\pm 1, \quad z=\pm 1
\end{eqnarray*}
which constrain the range of values for the coordinates $x$ and $z$. The 
zeros for $G(y) $ depends on the value of the cosmological constant 
$\Lambda$, which can be divided into the following
cases:
\begin{enumerate}
\item for $\Lambda>0$ (dS case), $G(y) $ has two zeros
  \begin{eqnarray*}
   y=\pm y_0, \quad y_0=\sqrt{1+\frac{\Lambda}{6\alpha^2}}>1;
  \end{eqnarray*}

\item for $\Lambda=0$ (Minkowski case), $G(y) $ has two zeros
$y=\pm 1$;

\item for $\Lambda<0$ (AdS case), the situation is a little more
complicated:
  \begin{itemize}
     \item if $-6\alpha^2<\Lambda<0$, $G(y) $ has two zeros
     \begin{eqnarray*}
     y=\pm y_0, \quad y_0=\sqrt{1+\frac{\Lambda}{6\alpha^2}}<1;
     \end{eqnarray*}
     \item if $\Lambda=-6\alpha^2$, $G(y) $ has a double zero $y=0$;
     \item if $\Lambda<-6\alpha^2$, $G(y) $ has no zeros.
  \end{itemize}
\end{enumerate}
Since $y$ plays the role of radial coordinate, its range is not restricted by 
the zeros of $G(y)$. However, the overall conformal factor in the
metric implies that $x+y=0$ is the conformal infinity, and in general it
suffices to consider the spacetime located on a single side of the conformal
infinity. Without loss of generality, we make the choice $x+y \geq 0$. Thus 
we get a variable range of $y$ depending on the value of $x$,
\begin{eqnarray*}
  y \in [-x, \infty).
\end{eqnarray*}
Putting together,  the coordinate ranges of the metric (\ref{metric1}) are 
given as follows,
\begin{eqnarray*}
  &t&\in(-\infty, \infty),\quad
  y\in[-x, \infty),\quad
  x\in[-1, 1],\\
  &z&\in[-1, 1],\qquad
  \phi\in[0, 2\pi).
\end{eqnarray*}

What is the role of zeros of the function $G(y)$ listed above? It will be 
shown later that many but not all of these zeros 
correspond to acceleration horizons. A given zero 
$y'$ (which can take either one of the values 
$\pm y_{0}$) of $G(y)$ corresponds to a horizon if and only if it is  
located inside the physical range $[-x,\infty)$ of $y$, i.e. 
it satisfies the inequality $x+y'>0$. All $y'$s whose values violate the 
above inequality are not horizons.

\section{Causal structures} \label{sec3}

\subsection{\texorpdfstring{$\Lambda>0$}{Lambda>0}} \label{3.1}

The procedure to draw the Carter-Penrose diagram is as follows. First, for convenience, we change the coordinates $x$ and $z$ into $\theta_{1}$ and $\theta_{2}$ via
\[
\theta_1 = \arccos(-x), \quad \theta_2 =\arccos(z)
\]
so that the angular part of the metric looks like that of a 3-sphere, 
\begin{eqnarray}
  \mathrm{d} \Omega_3^{2} = \mathrm{d} \theta_1^2 + \sin^2\theta_1 (\mathrm{d} \theta_2^2 + \sin^2\theta_2 \mathrm{d}
  \phi^2 ). \label{omega3}
\end{eqnarray}
Then we introduce the Eddington-Finkelstein coordinates,
\begin{eqnarray}
  u=t-y^*, \quad v=t+y^*,
\label{E-F}
\end{eqnarray}
where the tortoise coordinate $y^*$ is defined as
\begin{eqnarray}
  y^*=\int G^{-1} \mathrm{d} y
  =\frac{1}{2y_0}\log  \left|\frac{y-y_0}{y+y_0}\right|,\label{tortoise}
\end{eqnarray}
and both $u$ and $v$ belong to the range $(-\infty, \infty) $. In
this coordinate the metric becomes
\begin{eqnarray}
  \mathrm{d} s^2 = \frac{r^2}{\alpha^2}\left[-G(y) \mathrm{d} u
  \mathrm{d} v + \mathrm{d}\Omega_{3}^{2} \right], \quad r=(x+y)^{-1}.
\label{metric3}
\end{eqnarray} 

The Kruskal coordinates are introduced as
\begin{eqnarray*}
\tilde{u}=\pm\exp\left(-y_0u\right),\quad\tilde{v}=\pm\exp\left(y_0v\right),
\end{eqnarray*}
where $\tilde{u}$ and $\tilde{v}$ takes the same sign if $-x<y<y_0$, 
and they take opposite signs if $y\geq y_0$.
So there are totally 4 different combinations, 
each of which corresponds to a causal patch in the conformal 
diagrams to be drawn below. In each cases, one finds that 
\begin{eqnarray}
  \tilde{u}\tilde{v}=-\frac{y-y_0}{y+y_0},
\end{eqnarray}
and eq.(\ref{metric3}) becomes
\begin{eqnarray}
    \mathrm{d} s^2 = \frac{r^2}{\alpha^2}
    \left[-\frac{(y+y_0) ^2}{y_0^2}\mathrm{d} \tilde{u} \mathrm{d}
    \tilde{v} + \mathrm{d}\Omega_{3}^{2} \right],
\end{eqnarray}
where $y$ and $r$ are to be regarded as functions of
$\tilde{u}$ and $\tilde{v}$,
\begin{eqnarray*}
  y&=&y_0\frac{1-\tilde{u}\tilde{v}}{1+\tilde{u}\tilde{v}},\\  r&=&
  \frac{1+\tilde{u}\tilde{v}}{(y_0+x) -\tilde{u}\tilde{v}(y_0-x) }.
\end{eqnarray*}
Finally, the Carter-Penrose coordinates can be introduced by the
usual arctangent mappings of $\tilde{u}$ and $\tilde{v}$
\begin{eqnarray*}
  U&=&\arctan{\tilde{u}},\quad  V=\arctan{\tilde{v}},\\
  T&\equiv&U+V, \quad  R\equiv U-V,
\end{eqnarray*}
in terms of which the metric becomes
\begin{eqnarray}
 \mathrm{d} s^2 = \frac{1}{\alpha^2(y_0\cos{T}-\cos\theta_1
 \cos{R}) ^2}\left[- \mathrm{d} T^2 + \mathrm{d} R^2 +  
 \cos^2 (R) ~\mathrm{d} \Omega_3^{2} \right].
\end{eqnarray}

The values of the product
$\tilde{u}\tilde{v}$ at $y=y_0$, $r=0$ and $r=\infty$ are respectively 
\begin{eqnarray}
 \lim_{y \rightarrow y_0} \tilde{u}\tilde{v}=0,\quad
 \lim_{r \rightarrow 0} \tilde{u}\tilde{v}=-1,\quad \lim_{r \rightarrow
 \infty} \ \tilde{u} \tilde{v}=\frac{y_0+x}{y_0-x}.
 \label{product1}
\end{eqnarray}
These correspond to the various causal boundaries in the Carter-Penrose 
diagrams. From the coordinate transformations introduced above, it is 
easy to see that if the limit of $\tilde{u}\tilde{v}$ is 0 or $\infty$, the 
corresponding line is mapped into a null line; if the limit of 
$\tilde{u}\tilde{v}$ is $-1$, the corresponding line is mapped
into a spacelike line; if the limit is 1, 
the corresponding line is mapped into a timelike line. 
Since for $\Lambda>0$ we have $y_{0}>1$,  $\lim_{r \rightarrow
 \infty} \tilde{u} \tilde{v}$ has a finite positive
value varying with $x$. Therefore, the Carter-Penrose diagram in this case  
consists of the coordinate poles represented by the lines $r=0$, the 
horizon at $y=y_{0}$ and the (curved)  past and 
future conformal infinities at $r=\infty$. The corresponding diagram is 
depicted as in Fig.\ref{Fig1} (a). In this and all subsequent figures, lines 
labeled with $y=y_0$ or $y=-y_{0}$ represent acceleration horizons, 
$I^+$ and $I^-$ are respectively past and future infinities ($r=+\infty$
or $y=-x$), and $r=0$ or $I$ correspond to $y=+\infty$, the spacelike 
infinities.

Notice that the conformal infinities 
become exactly timelike (i.e. straight horizontal lines) at the particular 
value $x=0$. A similar Carter-Penrose diagram has been found for the 4D 
massless uncharged de Sitter C-metric in \cite{Dias:2003}.

\subsection{\texorpdfstring{$\Lambda=0$}{Lambda=0}}

\begin{figure}
\begin{center}
\includegraphics[width=\textwidth]{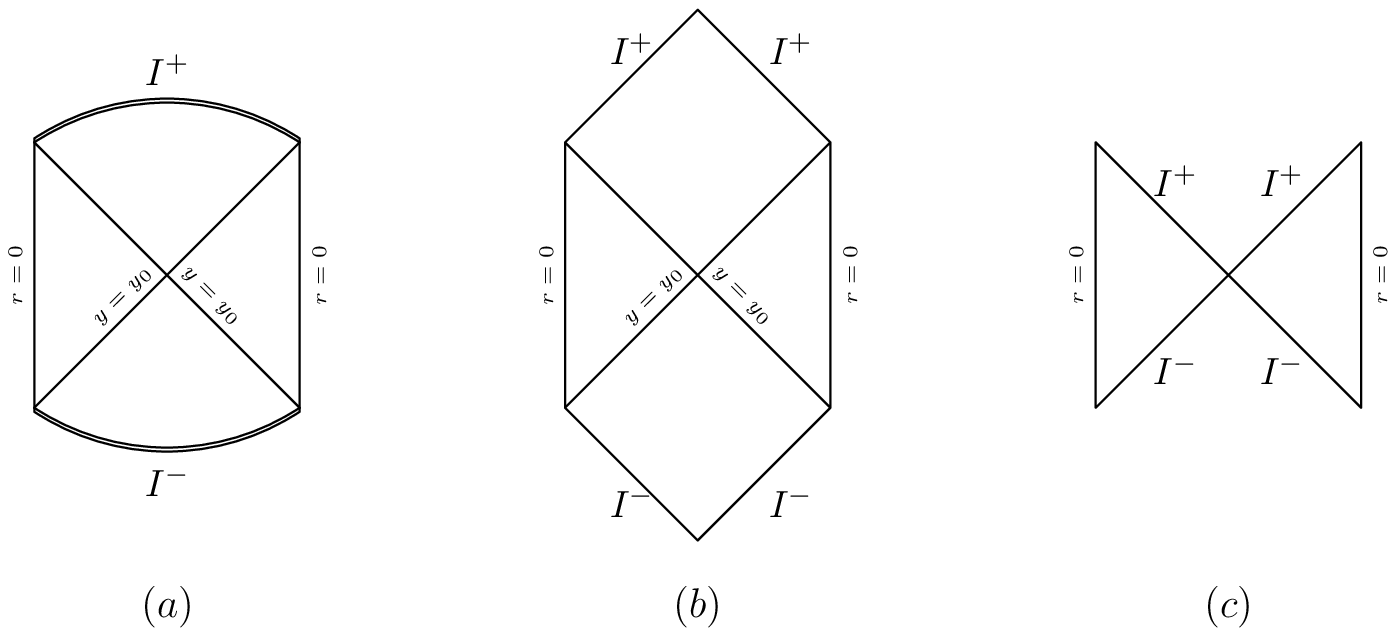}
\begin{minipage}{10.5cm}
\caption{Carter-Penrose diagrams for 
$\Lambda\geq 0$: (a) corresponds to both $\Lambda>0$ and the 
$-1<x<1$ case of $\Lambda=0$; (b) and (c) respectively corresponds to $
\Lambda=0$ with $x=1$ and $x=-1$.} \label{Fig1}
\end{minipage}
\end{center}
\end{figure}

The procedure for drawing Carter-Penrose diagrams for the case $\Lambda=0$ is very similar to the case $\Lambda>0$, the only difference lies in that we need to set $y_{0}$ to the specific value 1, and under this value of $y_{0}$ the analysis of $\lim_{r \rightarrow
 \infty} \tilde{u} \tilde{v}$ is a little bit more complicated:
\begin{itemize}
\item $-1<x<1$ : $y=1$ can be reached and is a horizon.  $\lim_{r 
\rightarrow\infty} \ \tilde{u} \tilde{v}$ is again finite, positive and 
varying with $x$. The Carter-Penrose diagram is the same as that for the $
\Lambda>0$ case, i.e. Fig.\ref{Fig1} (a);
\item $x=1$ : both the horizon at $y=1$ and the boundary at 
$y=-1$ can be reached, with the latter being null infinities. The 
Carter-Penrose diagram is depicted in Fig.\ref{Fig1} (b);
\item $x=-1$ : the minimum value of $y$ is $y=1$, which overlaps 
with null conformal infinities.  The Carter-Penrose diagram is depicted in 
Fig.\ref{Fig1} (c).
\end{itemize}

The same diagrams have appeared and were combined into the global 
visualization in \cite{Griffiths} which correspond to 
the massless uncharged C-metric without cosmological constant in 4D. 

\subsection{\texorpdfstring{$\Lambda<0$}{Lambda<0}}
For $\Lambda<0$ we have to analyze separately three subcases $-6\alpha^2<\Lambda<0$,
$\Lambda=-6\alpha^2$ and $\Lambda<-6\alpha^2$. Since the lower
bound for the coordinate $y$ depends on $x$, the 
Carter-Penrose diagrams will also change according to the value of $x$.

\subsubsection{\texorpdfstring{$-6\alpha^2<\Lambda<0$}{-6alpha^2< Lambda<0}}
 
This case corresponds to $0< y_{0} <1$ and the procedure to get the final conformal diagrams is basically the same as before. In the final step for analysing $\lim_{r \rightarrow
 \infty} \tilde{u} \tilde{v}$, we need to subdivide 
the range of $x$ into five subcases: $-1 \leq x<-y_0$, 
$x=-y_0$, $-y_0<x<y_0$, $x=y_0$ and $y_0<x<1$.

\begin{figure}[h,t]
\begin{center}
\includegraphics[width=\textwidth]{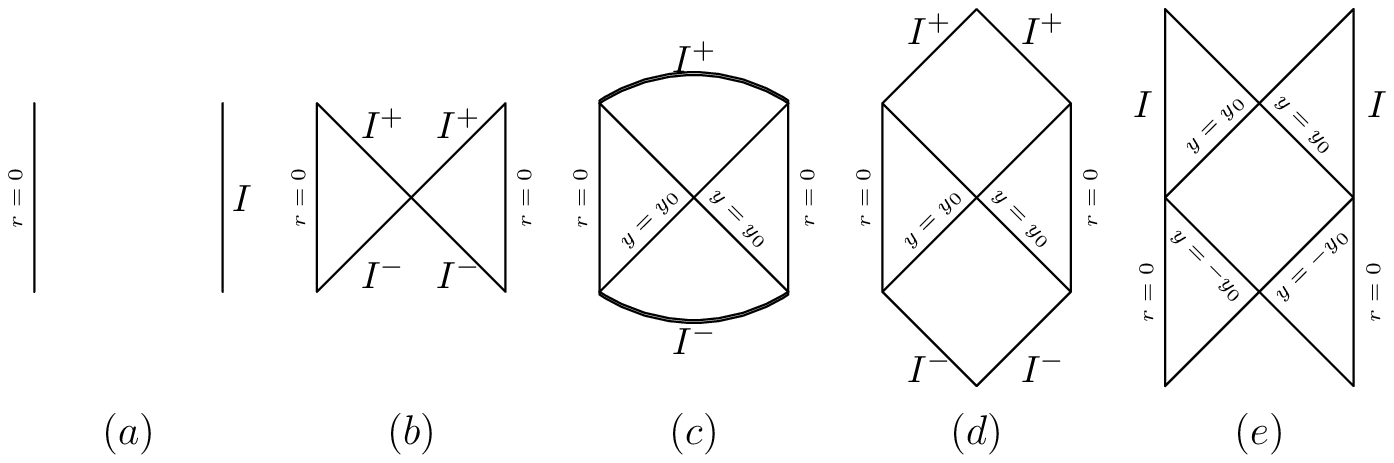}
\caption{Carter-Penrose diagrams for
$-6\alpha^2<\Lambda<0$} \label{Fig2}
\end{center}
\end{figure}

\begin{itemize}
\item $-1 \leq x<-y_0$: in this case  $\lim_{r \rightarrow \infty} 
\tilde{u} \tilde{v}$ takes a finite negative value which is varying with 
$x$. There is no horizons because $y_{0}$ is beyond the physical
region $[-x,\infty)$ for $y$. The corresponding causal diagram is 
depicted in Fig.\ref{Fig2} (a).

\item $x=-y_0$: $y=y_{0}$ overlaps with the conformal infinities at 
$r=\infty$, 
so there is no horizon and the Carter-Penrose diagram is 
shown in Fig.\ref{Fig2} (b).

\item $-y_0<x<y_0$ : $y=y_0$ is the only horizon because $-y_{0}$ is 
beyond the physical region of the coordinate $y$.  The Carter-Penrose 
diagram is similar to the $\Lambda>0$ case and is depicted in 
Fig.\ref{Fig2} (c).

\item $x=y_0$ : $y=y_0$ is the only horizon and $y=-y_{0}$ overlaps 
with the null conformal infinities. The Carter-Penrose 
diagram is shown in Fig.\ref{Fig2} (d).

\item $y_0<x<1$ : $y=y_0$ and $y=-y_0$ are both horizons and $\lim_{r 
\rightarrow \infty} \tilde{u} \tilde{v}$ takes a finite negative value which 
is varying with $x$. The Carter-Penrose diagram is shown in 
Fig.\ref{Fig2} (e). Note that for this case, the diagram is vertically not bounded -- there are repeating copies in the time direction.
\end{itemize}

\subsubsection{\texorpdfstring{$\Lambda=-6\alpha^2$}{Lambda=-6alpha^2}}
This case corresponds to $y_{0}=0$ and $G(y)=y^{2}$.  The tortoise 
coordinates given in (\ref{tortoise}) now cease to be meaningful, and we 
need to introduce $y^\ast$ in a $y_{0}$-independent way. Actually, inserting  $G(y)=y^{2}$ into the first equality of (\ref{tortoise}) naturally gives the correct $y^{\ast}$ for the present case,
\[
y^{\ast}=\int \frac{\mathrm{d}y}{y^{2}} = - \frac{1}{y}.
\]
So, introducing the sequence of coordinate changes
\begin{eqnarray*}
 u&=&t-y^\ast, \quad v=t+y^\ast,\\
 U&=&\arctan{u},\quad  V=\arctan{v},\\
 T&=&U+V, \quad  R= U-V,
\end{eqnarray*}
the metric can be rewritten as
\begin{eqnarray}
\mathrm{d}s^{2}=\frac{r^{2}}{\alpha^{2}\sin^{2}R}\left(
-\mathrm{d}T^{2}+\mathrm{d}R^{2}+\sin^{2}R \, \mathrm{d}\Omega_{3}^{2}\right).
\end{eqnarray}

\begin{figure}[h,t]
\begin{center}
\includegraphics[width=.9\textwidth]{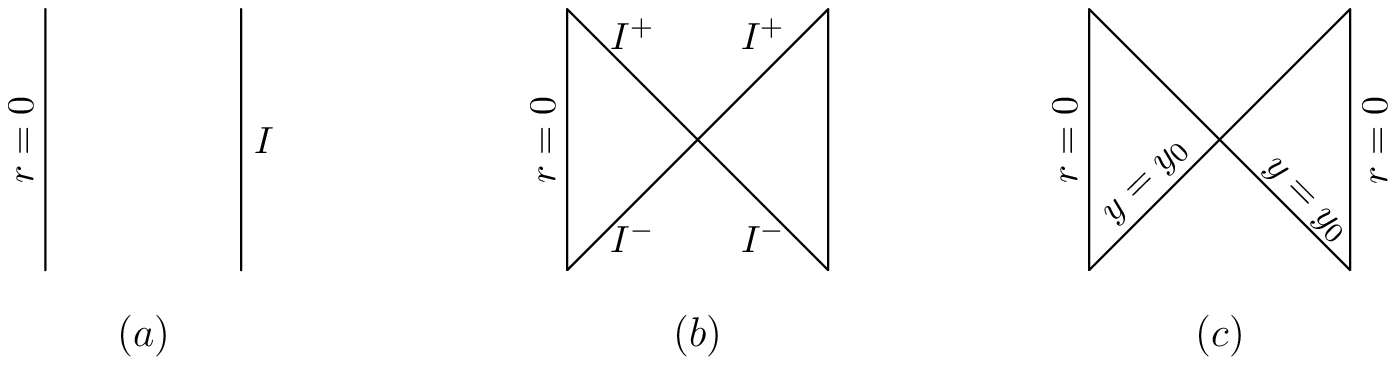}
\caption{Carter-Penrose diagrams for
$\Lambda=-6\alpha^2$} \label{Fig3}
\end{center}
\end{figure}

We need to subdivide the values 
of $x$ into 3 regions:
\begin{itemize}
\item $-1<x<0$ : No horizons exist and the Carter-Penrose diagram is 
depicted as in Fig.\ref{Fig3} (a);

\item $x=0$ : $y_{0}=0$ overlaps with the conformal infinities and hence 
no horizons exist and the Carter-Penrose diagram is depicted as in 
Fig.\ref{Fig3} (b).

\item $0<x<1$ : There is a double horizon at $y=0$ and the 
Carter-Penrose diagram is depicted as in Fig.\ref{Fig3} (c). 
\end{itemize}

\subsubsection{\texorpdfstring{$\Lambda<-6\alpha^2$}{Lambda<-6alpha^2}}

In this case we write $G(y)=y^2+y_{0}^{2}$, with $y_{0}^{2}=-1-
\frac{\Lambda}{6\alpha^{2}}>0$. Then the tortoise coordinate can be 
introduced as
\[
y^{\ast}=\int G^{-1}(y)\mathrm{d}y=\frac{1}{y_{0}}\arctan\frac{y}{
y_{0}}.
\]
Introducing the new coordinates $T=y_{0}t$, $R=y_{0}y^{\ast}$, the metric can be rewritten as
\[
\mathrm{d}s^{2}=\frac{r^{2}}{\alpha^{2}\cos^{2}R}
\left(-\mathrm{d}T^{2}+\mathrm{d}R^{2}+\cos^{2}R~
\mathrm{d}\Omega_{3}^{2}\right).
\]
Contrary to the cases $-6\alpha^2<\Lambda<0$ and 
$\Lambda=-6\alpha^2$ , the Carter-Penrose diagram of this case does not 
depend on the value of $x$ and contains no horizons. The corresponding 
diagram is depicted as in Fig.\ref{Fig3} (a).

In closing this subsection, let us remark that for each of the 
Carter-Penrose diagrams corresponding to $\Lambda<0$ listed above,  
one can find a 4D counter part for it in 
\cite{Dias:2002} respectively\footnote{For 4D massive C-metric with 
negative cosmological constant, the causal structure was also studied by Krtous in \cite{Krtous}, where the conformal diagrams are patched together to yield 3D global view.}.
Combining with the results obtained in previous subsections, we conclude 
that the Carter-Penrose diagrams for our metric with any value of 
$\Lambda$ are all analogous to the corresponding diagrams for the 4D 
massless unclarged C-metrics. The only difference is that
the geometry of each point in the diagrams are different (with one extra 
angular dimension in our case).  However, the same shapes
of the Carter-Penrose diagrams imply that our metric (\ref{metric1}) is the
5D generalization of the massless uncharged 4D C-metrics.

\section{Acceleration horizons}

In this section we shall show that every horizon appeared in Section 
\ref{sec3} is an acceleration horizon. 
This is achieved by use of some properly chosen coordinate 
transformations. In this process, the meaning of the parameter $\alpha$ 
will also gets clear: it is just the magnitude of the acceleration of the origin 
of the coordinate systems which we are going to choose. 

\subsection{\texorpdfstring{$\Lambda>0$}{Lambda>0}}
Let us first perform the following coordinate transformation
\cite{Griffiths:2001}
\begin{eqnarray}
  \tau &=& \alpha^{-1}\kappa t, \quad \quad \rho = \alpha^{-1}\kappa y^{-1}, \nonumber\\
  \theta_1 &=& \arccos (-x),   \quad \theta_2= \arccos z , \label{3.1-1}
\end{eqnarray}
with 
\begin{eqnarray}
\ell^2=6/\Lambda, \quad
\kappa=\sqrt{1+\alpha^2 \ell^2}. \label{Aell1}
\end{eqnarray}
Doing so the metric (\ref{metric1}) becomes
\begin{eqnarray}
  \mathrm{d} s^2 = \frac{1}{\gamma^2} \left[ -(1-\rho^2/\ell^2 )\mathrm{d} \tau^2 + \frac{\mathrm{d} \rho^2}{1-\rho^2/\ell^2} +\rho^2 \mathrm{d} \Omega_3^{2}  \right],
 \label{metricone}
 \end{eqnarray}
where
\begin{eqnarray}
  \gamma = \kappa - \alpha\rho\cos\theta_1.
  \label{3.1-2}
\end{eqnarray}

Now consider the timelike observer at fixed spacial position
$(\rho,\theta_{1},\theta_{2},\phi)$ in the spacetime described by the 
worldline $x^\mu(\lambda) = (\gamma \ell \lambda/
\sqrt{\ell^2-\rho^2}, \rho, \theta_{1},\theta_{2},\phi)$, where 
$\lambda$ is the proper time. The acceleration of the observer,
$a^\mu =  u^\nu\nabla_\nu u^\mu$, where $u^\mu = \mathrm{d} x^
\mu /\mathrm{d} \lambda$ is the proper velocity obeying $u^\mu 
u_\mu=-1$, has a magnitude
\begin{eqnarray}
  a^\mu a_\mu=\alpha^{2}
  + \frac {1}{\ell^{2} (\ell^{2} - \rho^{2} )} 
  \left( \kappa^{2}\rho ^{2} - 2\ell^{2}\alpha \kappa\rho\cos\theta_1
  +\ell^{2}\alpha^{2}\rho^{2}  \cos^{2}\theta_1\right). \label{mag}
\end{eqnarray}
We see that the observer at the origin $\rho=0$ (or
$y=\infty$), is being accelerated with a constant acceleration $|a|=
\alpha$. Notice that the choice $\alpha=0$ makes the metric 
(\ref{metricone}) become that of the usual dS spacetime in static 
spherical coordinates.
 Moreover, at $\rho=\ell$ (or $y=y_0$), the acceleration becomes
infinite which corresponds to the trajectory of a null ray. 
All observers held at $\rho=const$ see the null ray as an
acceleration horizon and they will never see events beyond this null
ray.

\subsection{\texorpdfstring{$\Lambda=0$}{Lambda=0}}

For $\Lambda=0$ we perform the following coordinate transformation
\begin{eqnarray}
  \tau &=& t, \quad  \rho = y^{-1},\nonumber\\
  \theta_1 &=& \arccos (-x),   \quad \theta_2= \arccos z.
\end{eqnarray}
The metric then becomes
\begin{eqnarray}
  \mathrm{d} s^2 = \frac{1}{\gamma^2} \left[ -(1-\rho^2 )\mathrm{d} \tau^2 + \frac{\mathrm{d} \rho^2}{1-\rho^2} +\rho^2 \mathrm{d} \Omega_3^{2}  \right],
 \label{metricq}
 \end{eqnarray}
where
\begin{eqnarray}
  \gamma = \alpha(1-\rho\cos\theta_1). \label{gamma0}
\end{eqnarray}
Consider the timelike observer at fixed spacial position described by
the worldline $x^\mu(\lambda) = (\gamma\lambda/\sqrt{1-\rho^2}, 
\rho, \theta_{1}, \theta_{2}, \phi)$. The proper acceleration
$a^\mu =(\nabla_\nu u^\mu) u^\nu$ now has a magnitude
\begin{eqnarray*}
  a^\mu a_\mu =\frac {\alpha ^{2}}{1-\rho^{2}}
  \left(1-\rho\cos\theta_{1}\right)^{2}.
\end{eqnarray*}
The observer at $\rho=0$ (or $y=\infty$) is being accelerated with 
a constant acceleration $|a|=\alpha$, while those at $\rho=1$ (or 
$y=y_0=1$) are being accelerated with infinite acceleration. 
This shows that for $\Lambda=0$, $y=y_{0}=1$ is indeed an accelerating 
horizon.

\subsection{\texorpdfstring{$\Lambda<0$}{Lambda<0}}

We need to consider 3 different cases.

\subsubsection{\texorpdfstring{$-6\alpha^2<\Lambda<0$}{-6alpha^2<Lambda<0}}
After performing the coordinate transformation of the same form with
(\ref{3.1-1}) but with 
\begin{eqnarray}
\ell^2=-6/\Lambda,\quad 
\kappa=\sqrt{\alpha^2 \ell^2-1},\label{Aell2}
\end{eqnarray}
we can rewrite the metric (\ref{metric1}) in the form (\ref{metricone}) 
with $\gamma$ taking the form (\ref{3.1-2}).
Then we consider the timelike observer $x^\mu(\lambda) = (\gamma \ell 
\lambda/\sqrt{\ell^2-\rho^2}, \rho,
\theta_1, \theta_2, \phi)$ at fixed spacial position. The magnitude of the 
proper acceleration observed by this observer takes exactly the same form 
as (\ref{mag}), now with $\ell$ and $\kappa$ given by (\ref{Aell2}).
The observer at $\rho=0$ (or $y=\infty$) is accelerated with a constant 
acceleration $|a|=\alpha$, and at $\rho=\ell$ (or $y=y_0$), the 
acceleration becomes infinite, signifying the existence of an acceleration 
horizon.

\subsubsection{\texorpdfstring{$\Lambda = -6\alpha^{2}$}
{Lambda=-6alpha^2}}

In this case we have $G(y)=y^{2}$, i.e. $y_{0}=0$. A possible
coordinate change is given by
\begin{eqnarray}
  \tau &=& \alpha^{-1}t, \quad  \rho =  \alpha^{-1} y^{-1},\nonumber\\
  \theta_1 &=& \arccos (-x),   \quad \theta_2= \arccos z,
\end{eqnarray}
after which the metric becomes
\begin{eqnarray}
  \mathrm{d} s^2 = \frac{1}{\gamma^2} \left( -\mathrm{d} \tau^2 + \mathrm{d} \rho^2 +\rho^2 \mathrm{d} \Omega_3^{2}  \right)
 \label{metric6alpha}
 \end{eqnarray}
with 
\begin{equation}
\gamma = 1-\alpha\rho\cos\theta_1. \label{gamma6alpha}
\end{equation} 
The metric (\ref{metric6alpha}) does not explicitly contain a horizon, and 
so is not applicable for evaluating infinite magnitude of the proper 
acceleration at the horizon. However, this form of the metric is explicitly 
conformal to Minkowski metric, as apposed to the previous 3 cases which 
are conformal to the standard form of de Sitter metric, so we keep it here. 

To actually realize that the double horizon at $y_{0}=0$ is indeed an 
acceleration horizon, we take an alternative route. We observe that the 
same metric (\ref{metricone}) can also describe the Einstein spacetime 
with $\Lambda=-6\alpha^{2}$, provided $\gamma$ is given in the form 
(\ref{3.1-2}) with 
\begin{eqnarray}
\kappa=\sqrt{2}\alpha\ell. \label{Aell3}
\end{eqnarray}
Therefore, we can borrow the 
result from the previous case (i.e. the case $-6\alpha^{2}<\Lambda<0$) 
and taking the limit $\kappa\rightarrow \sqrt{2}\alpha\ell$. In this way we 
should again find that the timelike observer at fixed spacial position will 
observe a proper acceleration of magnitude $a^\mu a_\mu=\infty$ at 
the horizon $\rho=\ell$ and $a^\mu a_\mu=\alpha^{2}$ at 
$\rho=0$.

\subsubsection{\texorpdfstring{$\Lambda<-6\alpha^2$}{Lambda<-6alpha^2}}

Now we make the coordinate transformation of the same form with
(\ref{3.1-1}) but with 
\begin{eqnarray}
\ell^2=-6/\Lambda,\quad
\kappa=\sqrt{1-\alpha^2\ell^2}.  \label{Aell4}
\end{eqnarray}
Doing so the metric (\ref{metric1}) is turned into the form 
\begin{eqnarray}
  \mathrm{d} s^2 = \frac{1}{\gamma^2} \left[ -(1+\rho^2/\ell^2 )\mathrm{d} \tau^2 + \frac{\mathrm{d} \rho^2}{1 +\rho^2/\ell^2} +\rho^2 \mathrm{d} \Omega_3^{2}  \right]
\label{metricthree}
\end{eqnarray} 
with $\gamma$ taking the form (\ref{3.1-2}).
Then the magnitude of the acceleration 
for the timelike observer $x^\mu(\lambda) = (\gamma \ell \lambda/\sqrt
{\ell^2+\rho^2}, \rho,\theta_1, \theta_2, \phi)$ at fixed spacial 
position turns out to be
\begin{eqnarray*}
  a^\mu a_\mu=\alpha^{2}+\frac {1}{\ell^{2}(\ell^{2} + \rho ^{2})}
  \left(\kappa^{2}\rho ^{2}+ 2\ell^{2}\alpha\kappa\rho \cos\theta_1 
  - \ell^{2}\alpha^{2}\rho ^{2}\cos^2\theta_1\right). 
\end{eqnarray*}
At $\rho=0$ (or $y=\infty$) the acceleration is again the same 
constant $|a|=\alpha$.  However,  the magnitude of the acceleration 
never blows up to infinity, showing that there is no acceleration horizons, 
which is in agreement with the fact that $G(y)$ has no zeros in the original 
metric in this case.

\subsection{Area of the horizons}

Having established that real zeros of $G(y)$ correspond to acceleration horizons, we now calculate the area of these horizons. The metric of the horizons can be written as
\[
\mathrm{d}s_{H}^{2}=\frac{\ell^{2}}{\gamma^{2}}\mathrm{d}\Omega_{3}^{2},
\]
where $\gamma$ is given by (\ref{3.1-2}) or (\ref{gamma0}) with $\rho=\ell$ according to different values of 
$\Lambda$.

The area of the horizons can be easily evaluated using the formula
\begin{eqnarray*}
  A=\int \mathrm{d} \theta_1
  \mathrm{d} \theta_2 \mathrm{d} \phi \sqrt{g_{H}},
\end{eqnarray*}
where the integration with respect to $\phi$ and $\theta_{2}$ is taken over 
$\phi\in [0, 2\pi)$, $\theta_{2} \in[0,\pi)$. The integration with respect to 
$\theta_{1}$ depends on the value of $\Lambda$: for $\Lambda \geq 0$, 
the integration is taken over $\theta_{1} \in [0,\pi)$, while for $ 
-6\alpha^{2} \leq\Lambda <0$, the integration is taken over $\theta_
{1}\in [0,\pi/2)$.

For $\Lambda>0$, we get a finite result
\[
A =2 \pi^2\left(\frac{6}{\Lambda}\right) ^{3/2}.
\]
For $-6\alpha^{2}\leq 
\Lambda\leq 0 $ cases, the area of the horizons always diverges. This 
result is in agreement with the common knowledge that the Einstein 
manifolds with $\Lambda >0$ are compact, while those with $\Lambda 
\leq 0$ are non-compact.

\section{Exterior geometry and relation to dS, AdS and Minkowski spacetimes}

By direct calculations it can be seen that the Weyl tensor for metric 
(\ref{metric1}) is identically zero. This signifies that the spacetime is 
conformally flat, very similar to the case of the standard form of dS, AdS 
and Minkowski spacetimes. In this section, we are aimed at exploring the 
relationship between our metric (\ref{metric1}) and the standard dS/AdS/
Minkowski spacetimes. It will be shown that our metric is indeed 
equivalent to the standard dS/AdS/Minkowski spacetimes, although 
written in different coordinate systems.
For this reason we adopt some exterior geometric techniques and consider 
the $\Lambda>0$ and $\Lambda<0$ cases as some hyperboloids 
embedded in 6D flat spacetime. For $\Lambda=0$, however, we shall show 
directly that our metric is the 5D Minkowski spacetime written in a 
particular coordinate system. 

\subsection{\texorpdfstring{$\Lambda>0$}{Lambda>0}}
The de Sitter spacetime can be represented as the 5-hyperboloid
\begin{eqnarray}
  -(X_0)^2 + (X_1)^2 + (X_2)^2 + (X_3)^2 + (X_4)^2 + (X_5)^2 = \ell^2
  \label{equationone}
\end{eqnarray}
embedded in the 6D Minkowski spacetime
\begin{eqnarray}
  \mathrm{d} s^2 = -(\mathrm{d} X_0)^2 + (\mathrm{d} X_1)^2 + (\mathrm{d} X_2)^2 + (\mathrm{d} X_3)^2 + (\mathrm{d} X_4)^2 + (\mathrm{d} X_5)^2
\label{metriconeone}
\end{eqnarray}
with $\ell^2=6/\Lambda$ and $\Lambda>0$.

Our metric for $\Lambda>0$, presented in the form 
(\ref{metricone}), fits nicely in the above 6D
picture if we parametrize the 5D hyperboloid (\ref{equationone}) with the  
following coordinate transformation
\begin{eqnarray*}
  X_0&=&\gamma^{-1}\sqrt{\ell^2-\rho^2}\sinh(\tau/\ell) , \quad  X_2=
  \gamma^{-1}\rho\sin\theta_1\sin\theta_2\sin\phi  ,\\
  X_1&=&\gamma^{-1}\sqrt{\ell^2-\rho^2}\cosh(\tau/\ell) , \quad  X_3=
  \gamma^{-1}\rho\sin\theta_1\sin\theta_2\cos\phi  ,\\
  X_5&=&\gamma^{-1}\left[-\kappa\rho\cos\theta_1+
  \alpha \ell^2\right] , \quad  X_4=\gamma^{-1}\rho\sin\theta_1\cos
  \theta_2,
\end{eqnarray*}
where $\rho$, $\tau$ and $\gamma$ given by (\ref{3.1-1}) and 
(\ref{3.1-2}) with $\ell$ and $\kappa$ given by (\ref{Aell1}). Similar transformations can be found in 
\cite{Griffiths:2001}, while treating the massless uncharged limit of the 4D 
dS C-metric. This justifies that the metric (\ref{metricone}) (and hence 
(\ref{metric1}) for $\Lambda>0$) is just the 5D de Sitter spacetime in a 
particular accelerating coordinate system.

\subsection{\texorpdfstring{$\Lambda=0$}{Lambda=0}}

We will show that for this particular value of $\Lambda$, 
our metric is just the 5D Minkowski spacetime written in a particular 
coordinate system. For this purpose, we will not resort to embedding into 
6D spacetime but stick to 5D description instead. The procedure to be 
carried out below is inspired by the work of \cite{Frolov} and the logic is 
basically the same as in the $\Lambda=0$ case of \cite{Liu Zhao}.

Consider the following 4D algebraic surface embedded in a 5D
Euclidean space,
\begin{eqnarray}
X_1^2+X_2^2+X_3^2+\left(\sqrt{X_4^2+X_5^2}-a\right)^2=b^2.
\label{embed}
\end{eqnarray}
For constants $a>b$, this equation describes a compact
4-dimensional surface of topology
$S^3\times S^1$. In fact, the surface can be thought of as the result of
pulling the center of a 3-sphere of radius $b$ everywhere around a circle
of radius $a$ lying in different dimensions.

We can parametrize the above surface in 5D Euclidean
space as follows:
\begin{eqnarray}
X_1 &=& \frac{\alpha}{B} \sin\theta\sin\chi\cos\phi, \quad
X_2 = \frac{\alpha}{B} \sin\theta\sin\chi\sin\phi,\quad
X_3 = \frac{\alpha}{B} \sin\theta\cos\chi, \nonumber\\
X_4 &=& \frac{\alpha}{B} \sinh\eta\cos\psi,\quad
X_5 = \frac{\alpha}{B} \sinh\eta\sin\psi,\label{p5}
\end{eqnarray}
where
\begin{eqnarray*}
B &\equiv& \cosh\eta - \cos\theta,\\
\alpha &\equiv& \sqrt{a^2-b^2},
\end{eqnarray*}
provided $\eta$ takes the special value
\begin{eqnarray*}
\eta=\eta_0, \quad \cosh\eta_0 = \frac{a}{b}.
\end{eqnarray*}
For variable values of $\eta$,  (\ref{p5}) is just another parametrization of 
the 5D Euclidean space.

Making some further coordinate transform
\begin{eqnarray*}
x&=&-\cos\theta,\quad
y=\cosh\eta,\quad
z=\cos\chi,
\end{eqnarray*}
it can be checked that the 5-dimensional Euclidean metric
\begin{eqnarray*}
\mathrm{d}s^2= \sum_{i=1}^5 dX_i^2
\end{eqnarray*}
is equivalent to the Wick rotated version of the metric (\ref{metric1}) with
$\Lambda=0$, i.e.
\begin{eqnarray}
\mathrm{d} s^2 &=& \frac{1}{\alpha^2(x+y) ^2}\left[(y^2-1) \mathrm
{d} \psi^2 + \frac{\mathrm{d} y^2}{y^2-1} \right.\nonumber\\
& &\qquad \left. + \frac{\mathrm{d} x^2}{1-x^{2}} + (1-x^{2}) \left
(\frac{\mathrm{d} z^2}{1-z^{2}} + (1-z^{2}) \mathrm{d} \phi^2
\right) \right].
\label{metric1p}
\end{eqnarray}
To actually obtain (\ref{metric1}), we need to make a Wick rotation $\psi
\rightarrow it$, which is equivalent to Wick rotating $X_{5}$ in the above.
This amounts to changing the geometry of the constant $y$ slices of the
spacetime to the embedding equation
\begin{eqnarray*}
X_1^2+X_2^2+X_3^2+\left(\sqrt{X_4^2-X_5^2}-a\right)^2=b^2
\end{eqnarray*}
in a 5D Minkowski spacetime. The above equation simply
describes the result of pulling the center of a 3-sphere everywhere along a
pair of hyperbolas.

\subsection{\texorpdfstring{$\Lambda<0$}{Lambda<0}}

The AdS spacetime can be represented as the 5-hyperboloid
\begin{eqnarray}
  -(X_0)^2 + (X_1)^2 + (X_2)^2 + (X_3)^2 + (X_4)^2 -(X_5)^2 = - \ell^2
  \label{equationtwo}
\end{eqnarray}
embedded in the 6D spacetime 
\begin{eqnarray}
  \mathrm{d} s^2 = -(\mathrm{d} X_0)^2 + (\mathrm{d} X_1)^2 + (\mathrm{d} X_2)^2 + (\mathrm{d} X_3)^2 + (\mathrm{d} X_4)^2 - (\mathrm{d} X_5)^2
\label{metrictwotwo}
\end{eqnarray}
with $\ell^2=-6/\Lambda$ and $\Lambda<0$. For different values $-6\alpha^2<\Lambda<0$, 
$\Lambda=-6\alpha^2$ and $\Lambda<-6\alpha^2$, we need different parametrizations for (\ref{equationtwo}).

\subsubsection{\texorpdfstring{$-6\alpha^2<\Lambda<0$}{-6alpha^2<Lambda<0}}

In this case, our metric can be written as in (\ref{metricone}). This metric  
follows from (\ref{equationtwo}) and (\ref{metrictwotwo})  if we 
parametrize  (\ref{equationtwo}) with the following coordinate 
transformation
\begin{eqnarray*}
  X_0&=&\gamma^{-1}\sqrt{\ell^2-\rho^2}\sinh(\tau/\ell) , \quad  X_2=\gamma^{-1}\rho\sin\theta_1\sin\theta_2\sin\phi  ,\\
  X_1&=&\gamma^{-1}\sqrt{\ell^2-\rho^2}\cosh(\tau/\ell) , \quad  X_3=\gamma^{-1}\rho\sin\theta_1\sin\theta_2\cos\phi  ,\\
  X_5&=&\gamma^{-1}\left[-\kappa\rho\cos\theta_1+\alpha \ell^2\right] , \quad  X_4=\gamma^{-1}\rho\sin\theta_1\cos\theta_2.
\end{eqnarray*}
with $\rho$, $\tau$ and $\gamma$ given by (\ref{3.1-1}) and 
(\ref{3.1-2}) with $\ell$ and $\kappa$ given in (\ref{Aell2}). Similar transformations can be found in 
\cite{Dias:2002}, while treating the massless uncharged limit of the 4D 
AdS C-metric. This justifies that the metric (\ref{metric1}) with $-6\alpha^{2}< \Lambda<0$ is just the 5D 
AdS spacetime written in a particular accelerating coordinate system.

\subsubsection{\texorpdfstring{$\Lambda=-6\alpha^2$}
{Lambda=-6alpha^2}}

In this case, our metric can be written as in (\ref{metric6alpha}). This 
metric follows from (\ref{equationtwo}) and (\ref{metrictwotwo}) if we introduce the following 
coordinate transformation
\begin{eqnarray*}
  X_0&=&\eta^{-1}\tau , \qquad\qquad
  X_2=\eta^{-1}\rho\sin\theta_1\sin\theta_2\sin\phi  ,\\
  X_1&=&\frac{1}{2}\eta^{-1}\left[1-(\rho^2\sin^2\theta_1
  +\ell^2\eta^2-\tau^2)\right] , \quad  
  X_3=\eta^{-1}\rho\sin\theta_1\sin\theta_2\cos\phi  ,\\
  X_5&=&\frac{1}{2}\eta^{-1}\left[1+(\rho^2\sin^2\theta_1
  +\ell^2\eta^2-\tau^2)\right] , \quad  
  X_4=\eta^{-1}\rho\sin\theta_1\cos\theta_2,
\end{eqnarray*}
where $\eta=\gamma/\alpha$ and $\gamma$ is given by (\ref{gamma6alpha}). 

\subsubsection{\texorpdfstring{$\Lambda<-6\alpha^2$}
{Lambda<-6\alpha^2}}

In this case, our metric takes the form (\ref{metricthree}), and it also 
follows from  (\ref{equationtwo}) and (\ref{metrictwotwo})  if we 
parametrize  (\ref{equationtwo}) with the 
following coordinate transformation 
\begin{eqnarray*}
  X_0&=&\gamma^{-1}\sqrt{\ell^2+\rho^2}\sinh(\tau/\ell) , \quad  
  X_2=\gamma^{-1}\rho\sin\theta_1\sin\theta_2\sin\phi  ,\\
  X_5&=&\gamma^{-1}\sqrt{\ell^2+\rho^2}\cosh(\tau/\ell) , \quad  
  X_3=\gamma^{-1}\rho\sin\theta_1\sin\theta_2\cos\phi  ,\\
  X_1&=&\gamma^{-1}\left[-\kappa\rho\cos\theta_1
  -\alpha \ell^2\right] , \quad 
   X_4=\gamma^{-1}\rho\sin\theta_1\cos\theta_2
\end{eqnarray*}
with $\rho$, $\tau$ and $\gamma$ given by (\ref{3.1-1}) and 
(\ref{3.1-2}) with $\ell$ and $\kappa$ given in (\ref{Aell4}). Similar transformations can be found in 
\cite{Podolsky:2002}, while treating the massless uncharged limit of 
the 4D AdS C-metric. This justifies that our metric (\ref{metricthree}) 
(hence (\ref{metric1}) with $\Lambda <-6\alpha^{2}$) is also an AdS 
spacetime written in a particular accelerating coordinate system.

Summarizing the above procedure, we conclude that the metric
(\ref{metric1}) is equivalent to either the standard dS or AdS spacetimes 
depending on the value of $\Lambda$. The same properties are shared by 
the weak field limit of the 4D C-metric as shown in detail in 
\cite{Dias:2002} and \cite{Dias:2003} respectively.

\section{4D interpretation}
The 5D metric (\ref{metric1}) admits a nice 4D
interpretation following the same fashion as we did in \cite{Liu
Zhao} for another metric. The basic strategy is to perform a boost and then
make a Kaluza-Klein reduction.

The boost is now made in the $t,\phi$ directions,
\begin{eqnarray*}
t &\rightarrow& T=t\cosh \beta - \phi \sinh \beta,\\
\phi &\rightarrow& \Phi=-t\sinh \beta + \phi \cosh \beta.
\end{eqnarray*}
Inserting the above into the metric (\ref{metric1}), we get
\begin{eqnarray*}
\mathrm{d} \tilde{s}_5^2 &=& \frac{1}{\alpha ^2(x+y) ^2}
\left [- \frac{G(y) -k^2 F(x) H(z) }{1-k^2}{\mathrm{d} T}^2 +\frac
{{\mathrm{d} y}^2}{G(y) }+\frac{{dx}^2}{F(x) }+\frac{F(x) }{H(z) }
{\mathrm{d} z}^2  \right.\\
 &+& \left. \frac{F(x) H(z) -k^2 G(y) }{1-k^2}{\mathrm{d} \Phi}^2 +
\frac{2k(F(x) H(z) -G(y) ) }{1-k^2} \mathrm{d} T \mathrm{d} \Phi
\right],
\end{eqnarray*}
where the boost velocity $k=\tanh \beta$ is used.

Now making a KK
reduction along the $\Phi$ axis using the formula
\begin{eqnarray*}
\mathrm{d} \tilde{s}_5^2 = e^{\varphi/\sqrt{3}}\mathrm{d}
\tilde{s}_4^{2} + e^{-2\varphi/\sqrt{3}}(\mathrm{d} \Phi +\mathcal{A}) ^2.
\end{eqnarray*}
we get the reduced 4D metric
\begin{eqnarray*}
\mathrm{d} \tilde{s}_4^2&=&\frac{1}{\alpha ^3(x+y) ^3}
\left(\frac{F(x) H(z) -k^2 G(y) }{1-k^2}\right) ^{1/2} \nonumber\\
&\times& \left[- \frac{F(x) G(y) H(z) (1 -k^2)  }{F(x) H(z) -k^{2}G(y)}
{\mathrm{d} T}^2+\frac{{\mathrm{d} y}^2}{G(y) }
+\frac{{\mathrm{d} x}^2}{F(x) }+F(x)
\mathrm{d} \theta_2^2 \right],
\end{eqnarray*}
where $z=\cos {\theta_2}$, together with the 4D Maxwell potential
\begin{eqnarray*}
\mathcal{A}=\frac{k[  F(x) H(z) -G(y) ]}{ F(x) H(z) -k^2 G(y) }
\mathrm{d} T,
\end{eqnarray*}
and the 4D Liouville field
\begin{eqnarray*}
e^{-2\varphi/\sqrt{3}}= \frac{1}{\alpha ^2(x+y) ^2}
\frac{F(x) H(z) -k^2 G(y) }{1-k^2}.
\end{eqnarray*}

The above reduction can also be realized on the level of classical actions.
Before the reduction, the action for the 5D vacuum Einstein
equation $R_{MN}-\frac{1}{2}g_{MN}R + \Lambda g_{MN}=0$ is
\begin{eqnarray}
S_5 = \int \mathrm{d}^5 x\sqrt{-g_{(5) }} \left( R_{(5) } - \Lambda
\right),
\label{action}
\end{eqnarray}
up to a possible boundary counter term which we omit. After the KK
reduction, the action becomes
\begin{eqnarray*}
{S}_4 = \int \mathrm{d}^4 x \sqrt{-g_{(4) }} \left( R_{(4) } - \frac{1}
{2}\left(  \partial \varphi \right) ^{2} - \Lambda e^{\varphi/\sqrt{3}} -
\frac{1}{4} e^{\varphi/\sqrt{3}} F_{\mu\nu}F^{\mu\nu}\right),
\end{eqnarray*}
where
\begin{eqnarray*}
F = F_{\mu\nu} \mathrm{d} x^\mu \wedge \mathrm{d} x^\nu \equiv
\mathrm{d} \mathcal{A}.
\end{eqnarray*}
We see that this is the action of Einstein-Maxwell-Liouville theory. At
$k=0$ the Maxwell field vanishes and only the 4D action reduces to that of
the Einstein-Liouville theory. On the other hand, keeping $k\neq 0$ while
setting $\Lambda=0$, we get a solution to the Einstein-Maxwell-dilaton
theory.

\section{Discussions}
In this article, we analyzed the general properties of the metric
(\ref {metric1}) found in \cite{Liu Zhao}. It turns out that this
metric resembles very much to the 4D massless uncharged C-metric. In
every cases of $\Lambda>0$, $\Lambda=0$ and $\Lambda<0$, the Carter
Penrose diagrams for the metric (\ref{metric1}) were found to have
the same shape as its 4D analogue found in \cite{Griffiths},
\cite{Dias:2003} and \cite{Dias:2002}. However, the appearance of a
fifth dimension gives room for Kaluza-Klein reduction, which leads
to a 4-dimensional interpretation in terms of
Einstein-Maxwell-Liouville theory.

Potentially, the 5D metric studied here might be useful in finding 
5D metrics with accelerating black holes inside and
constructing black ring solutions in 6 dimensions. 
In the latter respect, the usual 4D C-metric has
played a similar role in 5D black ring solutions \cite{Emparan:2001}.
Whether the same program can be carried on in the presence of one extra
dimension remains to be investigated. We hope we can come back on
that subject later on.

\section*{Acknowledgment} 

This work is supported by the National  Natural Science Foundation of 
China (NSFC) through grant No.10875059. L.Z. would like to thank the 
organizer and participants of ``The advanced workshop on Dark Energy 
and Fundamental Theory'' supported by the Special Fund for Theoretical 
Physics from the National Natural Science Foundation of China with grant 
no: 10947203 for comments and discussions.

\providecommand{\href}[2]{#2}\begingroup%\raggedright
\begin
{thebibliography}{13}

\bibitem{Emparan:2008}
R.~Emparan and H.~S. Reall, ``Black Holes in Higher Dimensions,''
{arXiv}
{\bf hep-th} (Jan, 2008)  \href{http://www.arXiv.org/abs/0801.3471v1}
{{\tt 0801.3471v1}}.

\bibitem{Obers}
N.~A. Obers, ``Black Holes in Higher-Dimensional Gravity,'' Lect.
Notes Phys. 769, pp211-258, 2009 [{arXiv:} {\bf hep-th} (Feb, 2008)
\href{http://www.arXiv.org/abs/0802.0519v1}{{\tt 0802.0519v1}}].

\bibitem{Liu Zhao}
Liu Zhao and Bin Zhu, ``C-metric like vacuum with non-negative
cosmological constant in five dimensions,'' [{arXiv:}
{\bf hep-th} (Oct, 2009)  \href{http://www.arXiv.org/abs/hep-th/
0910.3358v1}{{\tt hep-th/0910.3358v1}}].

\bibitem{Witten}
J.~Ehlers and W.~Kundt, {\em Exact solutions of the gravitational
field equations}.
\newblock in ``Gravitation: an introduction to current research'',
Ed. by L. Witten, John Willy \& Sons, Inc., 1962.

\bibitem{Emparan:2001}
R.~Emparan and H.~S. Reall, ``Generalized Weyl Solutions,''
Phys.Rev. D65 (2002)  084025 [{arXiv:} {\bf hep-th}
(Jan, 2001)  \href{http://www.arXiv.org/abs/hep-th/0110258v2}
{{\tt hep-th/0110258v2}}].

\bibitem{Griffiths}
J.~B. Griffiths, P.~Krtous, and J.~Podolsky, ``Interpreting the
C-metric,''   Class. Quant. Grav. 23 (2006)  6745-6766
[{arXiv:} {\bf gr-qc} (Sep, 2006) \href{http://www.arXiv.org/abs/gr-qc/
0609056v1}{{\tt gr-qc/0609056v1}}].

\bibitem{Hong}
K.~Hong and E.~Teo, ``A new form of the C-metric,'' Class. Quant.
Grav. 20 (2003)  3269-3277 [{arXiv:} {\bf gr-qc}
(May, 2003)  \href{http://www.arXiv.org/abs/gr-qc/0305089v2}{{\tt
gr-qc/0305089v2}}].

\bibitem{Dias:2002}
O.~J.~C. Dias and J.~P.~S. Lemos, ``Pair of accelerated black holes
in an anti-de Sitter background: the AdS C-metric,'' Phys.Rev. D67
(2003)  064001 [{arXiv:} {\bf hep-th} (Mar, 2003)
\href{http://www.arXiv.org/abs/hep-th/0210065v3}{{\tt hep-th/
0210065v3}}].

\bibitem{Dias:2003}
O.~J.~C. Dias and J.~P.~S. Lemos, ``Pair of accelerated black holes
in a de  Sitter background: the dS C-metric,'' Phys. Rev. D67 (2003)
084018 [{arXiv:} {\bf hep-th} (Jan, 2003)
\href{http://www.arXiv.org/abs/hep-th/0301046v2}{{\tt hep-th/
0301046v2}}].

\bibitem{Podolsky-talk}
J.~Podolsky, M.~Ortaggio,  ``Robinson–Trautman spacetimes in higher 
dimensions'', Class. Quantum Grav. 23, 5785 (2006); 

\bibitem{Podolsky-talk2} M.~Ortaggio, J.~Podolsky and 
M.~Žofka, ``Robinson–Trautman spacetimes with an electromagnetic field in higher dimensions'', Class. Quantum Grav. 25, 025006 (2008)

\bibitem{Frolov}
V.~P. Frolov and R.~Goswami, ``Surface Geometry of 5D Black Holes
and Black Rings,'' Phys. Rev. D75: 124001, 2007 [{arXiv} {\bf gr-qc} (Dec, 2006) \href{http://www.arXiv.org/abs/gr-qc/0612033v2}
{{\tt gr-qc/0612033v2}}].

\bibitem{Krtous} P.~Krtous, ``Accelerated black holes in an anti-de Sitter 
universe'', Phys. Rev. D 72 (2005) 124019, [{arXiv:} {\bf gr-qc} (Oct., 
2005) \href{http://www.arXiv.org/abs/gr-qc/0510101}{{\tt gr-qc/
0510101}}].

\bibitem{Griffiths:2001}
J.~Podolsky and J.~B. Griffiths, ``Uniformly accelerating black
holes in a de Sitter universe,'' Phys. Rev. D. 63 (2001) 024066
  [{arXiv:} {\bf gr-qc} (Jan, 2001)
  \href{http://www.arXiv.org/abs/gr-qc/0010109v2}{{\tt gr-qc/0010109v2}}].

\bibitem{Podolsky:2002}
J.~Podolsky, ``Accelerating black holes in anti-de Sitter
universe,'' Czech. J. Phys. 52 (2002) 1-10
  [{arXiv:} {\bf gr-qc} (Feb, 2002)
  \href{http://www.arXiv.org/abs/gr-qc/0202033v1}{{\tt gr-qc/0202033v1}}].

\end{thebibliography}\endgroup

\end{document}